\documentclass[usenatbib,onecolumn]{mn2e}
\usepackage{graphicx}

\newcommand{\nth}{_{\rm nth}}
\newcommand{\lsc}{_{\rm lsc}}
\newcommand{\pe}{_{\rm pe}}
\newcommand{\st}{_{\rm st}}
\newcommand{\rad}{_{\rm r}}
\newcommand{\Xray}{_{\rm x}}
\newcommand{\rms}{_{\rm rms}}

\newcommand{\colhead}[1]{\multicolumn{1}{c}{#1}}
\newcommand{\tablenotemark}[1]{$^{\rm #1}$}
\newcommand{\tablenotetext}[2]{\noindent$^{\rm #1}$ #2}
\newcommand{\phn}{\phantom{1}}

\title[DSR from Pulsar Wind Nebulae]
  {Diffusive Synchrotron Radiation from Pulsar Wind Nebulae}

\author[Fleishman \& Bietenholz]{G. D. Fleishman$^1$
and M. F. Bietenholz$^2$ \\
$^1$National Radio Astronomy Observatory, Charlottesville, VA
22903 \\ Ioffe Institute for Physics and Technology, 194021
St.\ Petersburg, Russia\\
$^2$Department of Physics and Astronomy, York University, Toronto, M3J~1P3, Ontario, Canada\\
}

\date{Version 8g \today}

\begin{document}

\maketitle

\begin{abstract}

Diffusive Synchrotron Radiation (DSR) is produced by charged particles
as they random walk in a stochastic magnetic field.  The spectrum of
the radiation produced by particles in such fields differs
substantially from those of standard synchrotron emission because the
corresponding particle trajectories deviate significantly from
gyration in a regular field. The Larmor radius, therefore, is no
longer a good measure of the particle trajectory.  In this paper we
analyze a special DSR regime which arises as highly relativistic
electrons move through magnetic fields which have only random
structure on a wide range of spatial scales.  Such stochastic fields
arise in turbulent processes, and are likely present in pulsar wind
nebulae (PWNe). We show that DSR generated by a single population of
electrons can reproduce the observed broad-band spectra of PWNe from
the radio to the X-ray, in particular producing relatively flat
spectrum radio emission as is usually observed in PWNe.  DSR can
explain the existence of several break frequencies in the broad-band
emission spectrum without recourse to breaks in the energy spectrum of
the relativistic particles.  The shape of the radiation spectrum
depends on the spatial spectrum of the stochastic magnetic field.  The
implications of the presented DSR regime for PWN physics are
discussed.

\end{abstract}


\section{Introduction}

In many astrophysical objects, radiation is produced by
relativistic charged particles moving in magnetic fields.  This
radiation is called magneto-bremsstrahlung or synchrotron
radiation. The theory of synchrotron radiation in the case of
regular magnetic fields is well established
\citep[e.g.,][]{Ginzb_Syr_1965, Pacholczyk1970}. However, in many
astrophysical fields there is a stochastic, turbulent component,
which can often dominate the regular field.  Astrophysical
magnetic fields, then, will often be variable over a wide range of
spatial and temporal scales.

In order to calculate spectrum of synchrotron emission from a
volume encompassing regions of varying field strength and
orientation, it is necessary to average the microscopic
emission intensity over the different field strengths and
orientations (as well as over the range of particle energies which
may be present). A common approach to this problem is to simply
average the standard synchrotron formulae for a regular field over
the varying magnetic field. However, in general, this approach is
only correct if the field can be described as regular over the
volume large compared to a typical particle orbit.

In the case of a field which has structure in a volume small
compared to the average particle orbit, the particle paths deviate
significantly from regular gyration around the field lines, and
the standard formulae can lead to incorrect results, particularly
for the emitted spectrum.  For example, if a turbulent magnetic
field is composed of \emph{random}\/ waves, then the ensemble of
these waves will result in an incoherent superposition of random
forces which will produce a stochastic electron trajectory quite
different from the circular orbit in the regular field.
Therefore, the standard synchrotron formulae will be in
error in this case \citep{Nik_Tsyt_1979, Topt_Fl_1987, Fl_2005b}.
To correctly calculate the emission spectrum in such a case
requires not only averaging over the regions of different field
strength and orientation but also over the many possible particle
paths, since the microscopic nature of the particle path strongly
influences the spectrum. This problem is highly non-linear: in
addition to the electron path affecting the nature of the
emission, electrons are efficiently mirrored from regions of high
magnetic field, and thus spend a disproportionate time in the
regions of lower field. As we will elaborate below, it is quite
possible that the fields in astrophysical sources, in particular
in pulsar wind nebulae (PWNe), will in fact have structure on very
small scales.

To calculate the radiation from electrons moving in stochastic
fields, and which do not necessarily have circular orbits, we turn
to the theory of Diffusive Synchrotron Radiation (DSR), which
attempts to calculate the average emission over the many possible
particle paths. The theory was established some time ago
\citep{Topt_Fl_1987, Topt_etal_1987}, and was reviewed recently by
\cite{Fl_2005b}. However, despite the many astrophysical
situations where it would be applicable, from the sun and geospace
to extragalactic objects \citep{Fl_2005b}, it is not yet in
widespread use.

So far, two special cases of DSR as applied to astrophysical sources
have been particularly considered.  The first one is DSR generated in
the presence of only \emph{small-scale}\/ random inhomogeneities and its
application to prompt gamma-ray burst spectra \citep{Fl_2005a}. Note
that a limiting special case of the DSR, namely one calculated in
perturbative 1-d approximation, is also referred to as "jitter"
radiation in the gamma-ray burst literature \citep[for a review see,
e.g.,][]{Piran_2005}, which is essentially the same physical process
as DSR\@. The second one is DSR generated in a superposition of
\emph{regular} and \emph{small-scale random} magnetic fields, which is
possibly relevant for the interpretation of spatially resolved wide-band
spectra observed from some extragalactic jets \citep{Fl_2005c}.

In this paper, we focus on pulsar wind nebulae (PWNe).  A
magnetized and highly relativistic pulsar wind, which is shocked
near the pulsar, is thought to be the primary source of both the
nebular field and relativistic particle population.  Plasma
instabilities in relativistic shocks are thought to produce
structure in the magnetic field down at very small scales, down to
the skin depth \citep[e.g.,][]{Kazimura_1998, Piran_2005}.
It is therefore quite possible that the magnetic field in PWNe
have significant structure on very small spatial scales, and
consequently that the standard synchrotron formulae may not allow
accurate calculations of the emission spectrum.  In this paper, we
calculate the broad-band emission spectra from PWNe, which are
often difficult to understand using only standard synchrotron
theory, using DSR.

We consider here a special case of DSR where the magnetic field
represents an incoherent superposition of waves with different scales
and random phases and orientations, and having a power-law spectrum.
We note here that such a field, consisting of {\em only} random waves
and having no regular component, is almost certainly an
oversimplification.  Real astrophysical fields, including those in
PWNe, will often contain a combination of stochastic and regular
magnetic fields.  The case considered here, however, is an interesting
limiting case, with the opposite limiting case being that of only
regular field which is well known from the application of the standard
synchrotron formulae.  We note however, that the problem of
calculating the emission in a more general case is non-linear (as
mentioned above) and the emission spectrum cannot necessarily be
easily determined from the two limiting cases of regular field only
and random field only.

Nonetheless, as we will show, the very simple model of field we
consider here produces broadband spectra for pulsar wind nebulae which
in remarkable agreement with observations, with a minimum of
additional parameters.  Thus, even if a complete description of the
synchrotron emission will require more elaborate modelling than is
possible in this paper, the present success strongly suggests that DSR
is a promising avenue for future, more detailed work.

We finally note that alternatives to the standard synchrotron models
for PWNe were already suggested some time ago. For example,
\cite{Arons1972} considered a nonlinear inverse Compton radiation on a
strong coherent electro-magnetic wave created by the rotation of the
magnetized pulsar, and found that such a model predicts high levels of
circular polarization in the diffuse radio emission, in contradiction
with observations. In contrast, the DSR process assumes the presence
of many stochastic incoherent static magnetic inhomogeneities in the
nebula volume, which will give rise to very little (if any) circular
polarization in agreement with observations.

The plan of the remainder of this paper is as follows: We give a brief
general formulation of DSR in \S~\ref{sformula}, and then discuss in
particular DSR as produced by a distribution of relativistic electrons
with a power-law distribution in energy in \S~\ref{spowerlaw}.  We
discuss the implications for PWNe in \S~\ref{simplic}, and then
compare DSR spectra to the observed ones for different PWNe in
\S~\ref{scompare}, and finally, give our conclusions in
\S~\ref{sconclude}.

\section{General Formulation}
\label{sformula}

We give here a summary of full DSR theory for the radiation
produced by particles moving in stochastic magnetic fields.  As
mentioned, it was first introduced by \citet{Topt_Fl_1987}, but
reviewed and slightly reformulated to use more convenient notation
by \citet{Fl_2005b}, which formulation we follow here. For a truly
random field, which represents a superposition
\begin{equation}
\label{random_waves}
 {\bf B}({\bf r},t) = \sum {\bf B}({\bf k},\omega) e^{i{\bf kr} -
 \omega t +\varphi_{\bf k}}
\end{equation}
of the waves with random phases
$\varphi_\mathbf{k}$, where the sum $\sum$ means the
summation and/or integration over all available Fourier components
with various $\omega$ and $\mathbf{k}$, we can write
\begin{equation}
\label{random_waves_corr*}
 \left<B_{\alpha}({\bf
k},\omega)B_{\beta}^*({\bf k}',\omega')\right> =
K_{\alpha\beta}({\bf k},\omega)\delta({\bf k}-{\bf k}')
\delta(\omega-\omega')
\end{equation}
where the brackets $\left<...\right>$ denote averaging over the
random phases, so only the waves with exactly the same pairs of
$\mathbf{k}$ and $\omega$ are correlated.  Stated another way, the
presence of the delta-functions in equation
(\ref{random_waves_corr*}) indicates that any waves with
$\mathbf{k}$ and $\mathbf{k}+d\mathbf{k}$ are entirely
uncorrelated.  As was pointed out in \citet{Fl_2005b}, the
approximation of the standard synchrotron radiation formulae is
not applicable in the case of a magnetic field composed of
uncorrelated waves, independent of their wave length.

Although we do not know the detailed statistical properties of the
magnetic turbulence in astrophysical objects, we consider here the DSR
spectrum generated in the presence of the truly random magnetic field
whose structure can be described by a power-law spectrum.  We discuss
possible implications of this regime for the interpretation of
wide-band spectra of PWNe.

According to \citet{Topt_Fl_1987} and \citet{Fl_2005b}, if the
deviation of the particle trajectory from the straight line (due
to effect of the \emph{regular} magnetic field) is small compared
to the correlation length of the random field, then the radiation
spectrum takes the form
\begin{equation}
\label{I_gen_3}
 I_{\omega}  = \frac{8Q^2 q(\omega)}{3 \pi  c} \gamma^2 \left(1+\frac{\omega\pe^2
 \gamma^2}{\omega^2}\right)^{-1} \Phi_1(s,r) \; + $$$$
 \frac{Q^2 \omega}{4 \pi c \gamma^2} \left(1+\frac{\omega\pe^2
 \gamma^2}{\omega^2}\right) \Phi_2(s,r)
,
\end{equation}
where $\gamma=E/Mc^2$ is the Lorentz-factor of emitting particle,
$Q$ and $M$ are its charge and the mass ($Q=e$ and $M=m$ are the
charge and the mass of the electron in practice), $\omega\pe$ is
the plasma frequency, which enters via the dielectric permeability
$\varepsilon(\omega) \simeq 1-\omega\pe^2/\omega^2$, $c$ is the
speed of light, and $\Phi_1(s,r)$ and $\Phi_2(s,r)$ stand for the
integrals:
\begin{equation}
\label{Fi_1_def}
 \Phi_1(s,r)  = 24s^2 \; \hbox{Im}\int_0^{\infty}dt
 \exp(-2s_0t) \times  $$$$
 \left[\coth t \exp\left(-2rs_0^3(\coth t - \sinh^{-1}t -
 t/2)\right)-\frac{1}{t}\right]
 ,
\end{equation}
\begin{equation}
\label{Fi_2_def}
 \Phi_2(s,r)  = 4rs^2 \hbox{Re} \int_0^{\infty}dt \frac{\cosh t-1}{\sinh
 t}\times
 $$$$
  \exp\left(-2s_0t-2rs_0^3(\coth t - \sinh^{-1}t -
 t/2)\right)
 ,
\end{equation}
which depend on the dimensionless parameters $s_0$,  $s$, $r$:
\begin{equation}
\label{s_def}
 s_0=(1 -i)s  =  \frac{1-i}{8\gamma^2}
  \left(\frac{\omega}{q(\omega)}\right)^{1/2}\left(1+\frac{\omega\pe^2
 \gamma^2}{\omega^2}\right)
 ,
\end{equation}
\begin{equation}
\label{r_def}
 r =  32 \gamma^4
  \left(\frac{\omega_{B\bot}}{\omega}\right)^2\left(1+\frac{\omega\pe^2
 \gamma^2}{\omega^2}\right)^{-3}
,
\end{equation}
$q(\omega)$ is the rate of scattering of the particle by magnetic
inhomogeneities, which will be specified below, and
$\omega_{B\bot}=QB_{\bot}/(Mc)$, where $B_{\bot}$ is
the regular magnetic field component transverse to the line of
sight.

In essence, it would be highly desirable to consider the joint
effect of the random and the regular magnetic fields on the
emitted radiation since both components can coexist in the nebula
volume. Unfortunately, the full theory which allows for strong
electron path deflections due to {\em both}\/ random and
regular field components is currently unavailable.  Therefore, we
considered here a limiting special case opposite to widely
accepted case of synchrotron emission in the regular field.  In
particular, we consider the case when there is only random
magnetic field and no regular field, i.e., $B_{\bot} \rightarrow
0$, so equation (\ref{I_gen_3}) is applicable and can be
simplified further since $r \rightarrow 0$ and
\begin{equation}
\label{Fi_migdal_def*}
 \Phi_1(s,r=0)\equiv\Phi(s)  = 24s^2 \int_0^{\infty}dt
 \exp(-2st)\sin (2st)
  \times $$$$
 \left[\coth t -\frac{1}{t}\right]
 ,\ \quad \Phi_2(s,r=0)=0,
\end{equation}
thus
\begin{equation}
\label{I_DSR_stoch}
 I_{\omega}  = \frac{8Q^2 q(\omega)}{3 \pi  c} \gamma^2 \left(1+\frac{\omega\pe^2
 \gamma^2}{\omega^2}\right)^{-1} \Phi(s).
\end{equation}
Although the integration in equation (\ref{Fi_migdal_def*}) cannot
be performed analytically, the Migdal function $\Phi(s)$ has
simple asymptotes for small and large values of $s$
\citep{Migdal,Migdal_1956}:
\begin{equation}
\label{migdal_asympt}
 \Phi(s) \simeq 1, \hbox{if} \ s \gg 1,
 \qquad  \Phi(s) \simeq 6s, \hbox{if} \ s \ll 1.
\end{equation}

Let us estimate the spectrum of radiation from a single relativistic
particle with Lorentz-factor $\gamma$, moving in the stochastic field
consisting of a broad power-law spectrum of magnetic waves.

\begin{equation}
\label{power_spectr}
 K_{\alpha \beta}(\mathbf{k},
\omega)= \frac{A_{\nu}\delta(\omega -
\omega(\mathbf{k}))}{2(k_{0}^2+k^2)^{\nu/2+1}}\left(\delta_{\alpha
\beta} - {k_{\alpha} k_{\beta} \over k^2}\right),\
 $$$$
 A_{\nu}= {\Gamma(\nu/2+1) k_{0}^{\nu-1} \left<B\st^2\right> \over 3
 \pi^{3/2}\Gamma(\nu/2-1/2)},
\end{equation}
where $\Gamma(z)$ is the Euler gamma-function, $k_{0}=2\pi/L_{0}$,
$L_0$ is the largest scale of the magnetic turbulence, $\nu$  is
the spectral index of the turbulence,  $\left<B\st^2\right>$ is the
mean square of the random magnetic field. Here, in
equation (\ref{power_spectr}), we assume that the magnetic fluctuations
are composed of some propagating eigen-modes with a dispersion
relation $\omega = \omega(\mathbf{k})$, which results in the
$\delta$-function $\delta(\omega - \omega(\mathbf{k}))$ in equation
(\ref{power_spectr}). Therefore, the spatial dependence on
$\mathbf{k}$ and temporal dependence on $\omega$ are strictly
correlated for magnetic turbulence. Frequently, e.g., for
normal MHD waves, the group velocity $| {\bf v} | = |~\partial
\omega /\partial {\bf k}|$ is much less than the speed of light,
$c$. For such cases we can adopt $\omega(\mathbf{k})=0$, and,
accordingly, $\delta(\omega -
\omega(\mathbf{k}))\approx\delta(\omega )$, which is the
approximation of a quasi-static magnetic field, and therefore the
turbulence spectral index $\nu$ describes the distribution of the
magnetic energy over different spatial scales.

A key parameter in the DSR theory, the rate of the particle
scattering by magnetic inhomogeneities, $q(\omega)$, can
be approximated by \citep[see Eq.\ 35 in][]{Fl_2005b}:
\begin{equation}
\label{q_fin_appr}
 q(\omega)=
 \frac{\sqrt{\pi}\Gamma(\nu/2)\omega\st^2\omega_0^{\nu-1}}{3\Gamma(\nu/2-1/2) \gamma^2
 \left[(a\omega/2)^2\left(\gamma^{-2} + \omega\pe^2/\omega^2\right)^2+\omega_0^2\right]^{\nu/2}} ,
\end{equation}
in the case of magnetic field spectrum given by equation (\ref{power_spectr}),
and where $a$ is a number of the order of unity \citep{Fl_2005b}, and
$\omega_0= k_0c$, $\omega\st^2=Q^2\left<B\st^2\right>/(Mc)^2$ is
the mean square of the cyclotron frequency in the random magnetic field.

\begin{figure*}
\includegraphics[height=4in]{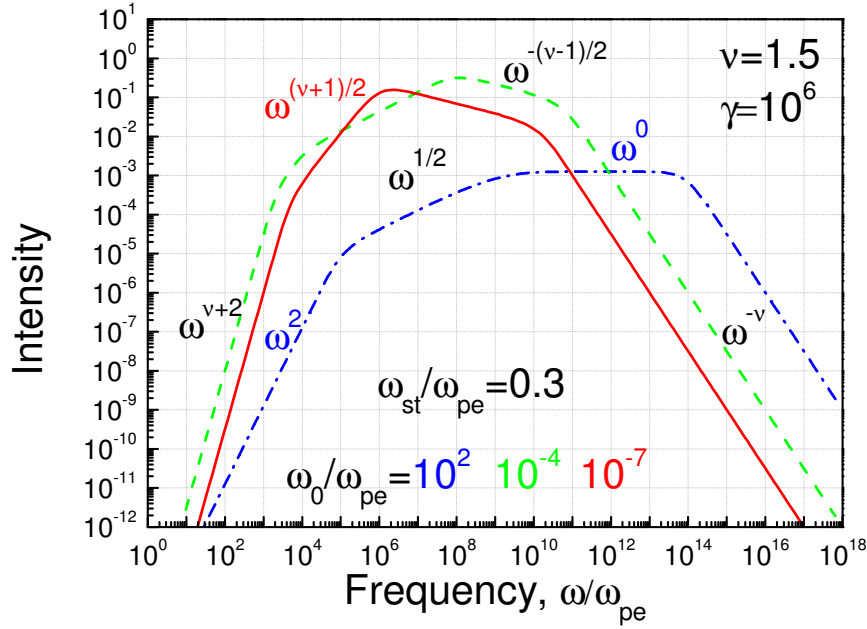}
\caption{Single particle DSR spectra for
$\nu=1.5$ and $\gamma=10^6$, where $\nu$ is the spectral index of
the turbulence, $\gamma$ the Lorentz factor of the particle, and
$\omega\pe$ the plasma frequency.  Cases of small-scale
($\omega_0/\omega\pe=10^2$) and large-scale
($\omega_0/\omega\pe=10^{-4}$ and $10^{-7}$) random magnetic field
are shown. The shape of the DSR radiation spectrum changes
significantly as the largest scale of the field ($L_0=2\pi
c/\omega_0$) changes as described in the text; in particular, the
regime of large-scale magnetic field (red/solid and green/dashed
curves) differs substantially from the regime of the small-scale
field (blue/dash-dotted curve) which is described in \citet{Fl_2005a}.
\label{pwn_single}}
\end{figure*}

In Figure~\ref{pwn_single} we show the DSR spectra calculated by
the numerical integration of equation (\ref{I_DSR_stoch}) in the
case of strong random magnetic field, i.e., under
condition
\begin{equation}
\label{large_scale}
 \omega\st \gg \omega_0
\end{equation}
along with the DSR spectrum in a small-scale field. Although
equation (\ref{I_DSR_stoch}) in general must be integrated numerically, we
can analytically obtain several characteristic asymptotes, which
give a good qualitative idea of the general shape of the DSR
spectrum.

As we will see, this regime contains a characteristic frequency
\begin{equation}
\label{w_lsc_def} \omega\lsc =\left[\left(\frac{2\pi
c}{L_0}\right)^{\nu-1}\frac{e^2\left<B\st^2\right>}{m^2
c^2}\right]^\frac{1}{\nu+1}=
 \left(\frac{\omega\st}{\omega_0} \right)^{\frac{2}{\nu+1}} \omega_{0} =
 \left(\frac{\omega_{0}}{\omega\st} \right)^{\frac{\nu-1}{\nu+1}} \omega\st
 ,
\end{equation}
 which plays a role similar to some extent to the
 role of the frequency $\omega_{B\bot}$ in the standard
 synchrotron theory.
At high frequencies
\begin{equation}
\label{high_freq}
 \omega\gg \omega\lsc \gamma^2 \equiv
 \left(\frac{\omega\st}{\omega_0} \right)^{\frac{2}{\nu+1}} \omega_{0} \gamma^2 \equiv
 \left(\frac{\omega_{0}}{\omega\st} \right)^{\frac{\nu-1}{\nu+1}} \omega\st \gamma^2
\end{equation}
we have $s \gg 1$, so the radiation spectrum has the standard
high-frequency form
\begin{equation}
\label{I_DSR_high}
 I_{\omega}  = \frac{8Q^2 q(\omega)}{3 \pi  c} \gamma^2  ,
\end{equation}
with the spectral asymptote $I_\omega \propto \omega^{-\nu}$,
typical for the high-frequency perturbative regime of DSR\@. Note
that if the magnetic field were regular with the same strength,
then the bounding frequency would be $\omega\st\gamma^2$ in place
of $\omega\lsc\gamma^2$ (equation \ref{high_freq}), and the
radiation intensity would decrease exponentially rather than as a
power-law (equation \ref{I_DSR_high}). The decrease of the
bounding frequency (compared with the regular field regime)
happens because the deviation of the particle trajectory from the
straight line occurs more slowly in the random than in regular
field, thus the region of applicability of the perturbation theory
(asymptote (\ref{I_DSR_high})) broadens towards lower frequencies.

Accordingly, the amount of radiated energy in the random
magnetic field is less than in the regular magnetic field with the
same energy density. Stated another way, larger random (than
regular) magnetic field is required to provide the same radiative
losses.

The parameter $s$ decreases with frequency, and when it falls
below unity at
\begin{equation}
\label{low_freq}
 \omega < \omega\lsc \gamma^2 ,
\end{equation}
the perturbation theory is no longer valid.  For $s \ll 1$ we have
$\Phi (s) \simeq 6s$, therefore
\begin{equation}
\label{I_DSR_nonper}
 I_{\omega}  = \frac{2Q^2 }{ \pi  c}(\omega q(\omega))^{1/2}  .
\end{equation}
This expression is valid down to relatively low frequencies, where
parameter $s$ increases again under the influence of the effect of
density (term $\omega\pe^2/\omega^2$) and again reaches the unity at a
sufficiently low frequency.

The asymptotic regime $\Phi (s) \simeq 6s$, at $s
\ll 1$,  is due to multiple scattering of the fast particle by the
uncorrelated long waves composing the random magnetic field at the
scales $l > c/\omega\lsc$. Even though the perturbation of the
particle trajectory due to any single Fourier-component of the
random field is small, their cumulative effect results in
significant angular diffusion of the charged particle.
Accordingly, the direction of the particle's motion changes by a value
exceeding the characteristic beaming angle of emission ($\vartheta
\sim \gamma^{-1}$), which leads to a suppression of the emission
compared with that predicted by the perturbation theory ($I_\omega
\propto \omega^{-\nu}$).

Note that this  essentially non-perturbative regime of DSR has no
direct analogies in other emission mechanisms. In particular, it
cannot be obtained by either perturbation theory or any averaging
of the standard synchrotron radiation, since the real particle
trajectory in the presence of the large-scale random magnetic
inhomogeneities deviates strongly from both the straight line and
a circle.

In the region of applicability of equation (\ref{I_DSR_nonper}), the
radiation spectrum is composed of two or three power-law
asymptotes depending on the relation between $\omega\pe\gamma$
and $\omega_{0}\gamma^2$. If $\omega_{0} \ll \omega\pe/\gamma$
then $\omega_{0}^2$ can be discarded in the denominator in
$q(\omega)$ (\ref{q_fin_appr}) at all frequencies. Accordingly,
discarding also the term $\omega\pe^2/\omega^2$, which is valid
at $\omega \gg \omega\pe\gamma$, we obtain
\begin{equation}
\label{I_DSR_medi}
 I_{\omega} \sim
  \frac{Q^2}{c}\omega\st\left(\frac{\omega_{0}\gamma^2}{\omega}\right)^{(\nu-1)/2}; \
  \omega\pe\gamma \ll \omega \ll \omega\lsc\gamma^2,
\end{equation}
where we have omitted a numeric coefficient near unity for
simplicity. It is important to note that for the typical
turbulence spectra with $\nu =$1 -- 2
\citep{turbulence,Goldreich_Sridhar_1995,Beresnyak+2005} the DSR
spectrum (equation \ref{I_DSR_medi}) is relatively flat having the
spectral index $\alpha = (\nu-1)/2 =$ 0 -- 0.5 (where $S_f \propto
f^{-\alpha}$).

In the other case, $\omega_{0} \gg \omega\pe/\gamma$, the lower bound
of this region shifts towards larger frequencies, and the spectrum has
the form given by (\ref{I_DSR_medi}) at $\omega_{0}\gamma^2 \ll \omega
\ll \omega\lsc\gamma^2$. At low frequencies, $\omega\pe^2/\omega_{0}
\ll \omega \ll \omega_{0}\gamma^2$, the scattering rate does not
depend on frequency, and $I_{\omega} \sim \omega^{1/2}$ as in the case
of small-scale magnetic inhomogeneities \citep{Fl_2005b}. Then, for
even lower frequencies, $\omega \ll \omega\pe^2/\omega_{0}$, (or, if
$\omega_{0} \ll \omega\pe/\gamma$, for $\omega \ll \omega\pe\gamma$),
the effect of density, described by the term $\omega\pe^2/\omega^2$,
dominates at the denominator of $q(\omega)$. Discarding other terms,
we obtain
\begin{equation}
\label{I_DSR_medi_low}
 I_{\omega} \sim
  \frac{Q^2}{c} \ \frac{\omega\st}{\gamma} \ \frac{\omega_{0}^{(\nu-1)/2}\omega^{(\nu+1)/2}}
  {\omega\pe^{\nu}}.
\end{equation}
Therefore, at these low frequencies the spectrum falls with
frequency as  $I_{\omega} \sim \omega^{(\nu+1)/2}= \omega^{1-1.5}$
for    $\nu  =$1 -- 2.  Finally, at a sufficiently low frequency
the parameter $s$ will again be larger than unity due to effect of
plasma dispersion (term  $\omega\pe^2/\omega^2$), so a standard
low-frequency DSR asymptote, $I_{\omega} \sim \omega^{\nu+2}$,
applies.

We conclude that the spectrum of electromagnetic emission produced
by a single relativistic particle in the presence of strong random
magnetic field is entirely different from the standard synchrotron
spectrum, and also deviates significantly from that in the
small-scale random magnetic field.

\section{Diffusive Synchrotron Radiation spectrum produced by
power-law electron distribution}
\label{spowerlaw}

Let us proceed now to the DSR spectra produced by a power-law
distribution of relativistic electrons
\begin{equation}
\label{distr_fun_power}
  dN_e(\gamma) = (\xi-1)N_e \gamma_1^{\xi-1} \gamma^{-\xi} d\gamma, \ \ \gamma_1 \le \gamma
  \le \gamma_2,
\end{equation}
where $N_e$  is the number density of electrons with
energies ${\cal E} \ge mc^2 \gamma_1$, $\xi$ is the power-law
index of the distribution, and $\gamma_1 \gg 1$.

Although several different regimes of DSR are possible,
depending on the value of $\xi$, we consider the case when
\begin{equation}
\label{main_power_case}
  \nu < \xi < 2\nu + 1,
\end{equation}
which is probably of the most practical importance. Indeed, the
turbulence spectral index is typically $\nu < 2$ (e.g., $\nu
\approx 1.7$ for the Kolmogorov turbulence), while the particle
index $\xi > 2$ (e.g., $\xi \approx 2.7$ for the galactic cosmic
rays).  In this regime, the standard non-thermal spectrum (typical
also for synchrotron radiation), $P_\omega \propto
\omega^{-\alpha\nth}$, where $\alpha\nth=(\xi-1)/2$, produced by
the inner part $\gamma_1 \ll \gamma \ll \gamma_2$ of the
distribution (equation~\ref{distr_fun_power}) is steeper than the
non-perturbative DSR spectrum, $P_\omega \propto
\omega^{-(\nu-1)/2}$, but shallower than the high-frequency
perturbative spectrum, $P_\omega \propto \omega^{-\nu}$.

\begin{figure*}
\includegraphics[height=4in]{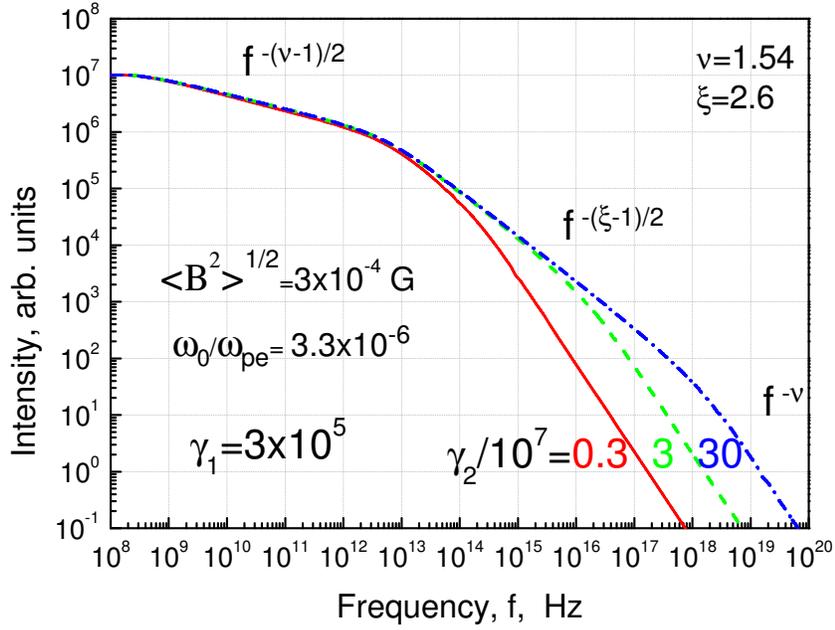}
\caption{Broad-band DSR spectra as produced
by three different power-law distributions
(equation~\ref{distr_fun_power}) of ultra-relativistic nebular
electrons, which extend from a Lorentz factor $\gamma_1=0.3\times
10^6$ to ones of $\gamma_2=3\times 10^6$, $3\times 10^7$, and
$3\times 10^8$, respectively.  The values of the model parameters
were chosen so as to match the spectral energy distribution of the
Crab Nebula (see text, \S~\ref{scrab}).  The values adopted for
the spectral index of the turbulence, $\nu$, the energy index of
the power-law electron distribution, $\xi$, the plasma frequency,
$\omega\pe$, and the mean square of the random magnetic field,
$\left<B\st^2\right>$, are indicated in the figure.
\label{DSR_Crab}}
\end{figure*}

The bulk of the DSR energy produced by a single particle is emitted at
frequencies $\omega \sim \omega\lsc \gamma^2$ (if $\nu < 3$).
Accordingly, it is easy to estimate that in the region of
``intermediate'' frequencies
\begin{equation}
\label{inner_freq_range}
   \omega\lsc\gamma_1^2 \ll \omega   \ll \omega\lsc \gamma_2^2,
\end{equation}
the standard well-known non-thermal (synchrotron-like) spectrum
$P_\omega \propto \omega^{-\alpha\nth}$ is formed. However, the
DSR spectrum will deviate significantly from the standard
synchrotron spectrum at high ($\omega > \omega\lsc \gamma_2^2$)
and low ($\omega < \omega\lsc \gamma_1^2$) frequencies, where it
will reproduce the single-particle spectra, $P_\omega \propto
\omega^{-\nu}$ (in place of the exponential synchrotron cut-off)
and $P_\omega \propto \omega^{-(\nu-1)/2}$ (in place of $P_\omega
\propto \omega^{1/3}$) respectively.  An example of the DSR
spectrum generated by the power-law electron distribution is given
in Figure~\ref{DSR_Crab}. We note that at sufficiently low
frequencies the spectrum decreases as $P_\omega \propto
\omega^{(\nu+1)/2}$ and then as $P_\omega \propto \omega^{\nu+2}$,
but these regions may or may not be observable from the Earth
depending on the source parameters. Below we assume that
this happens beyond the observed spectral range and, thus, do not
consider further the effect of plasma density on the DSR spectra.

\section{Implications for Pulsar Wind Nebulae}
\label{simplic}

\begin{figure*}
\includegraphics[height=4in]{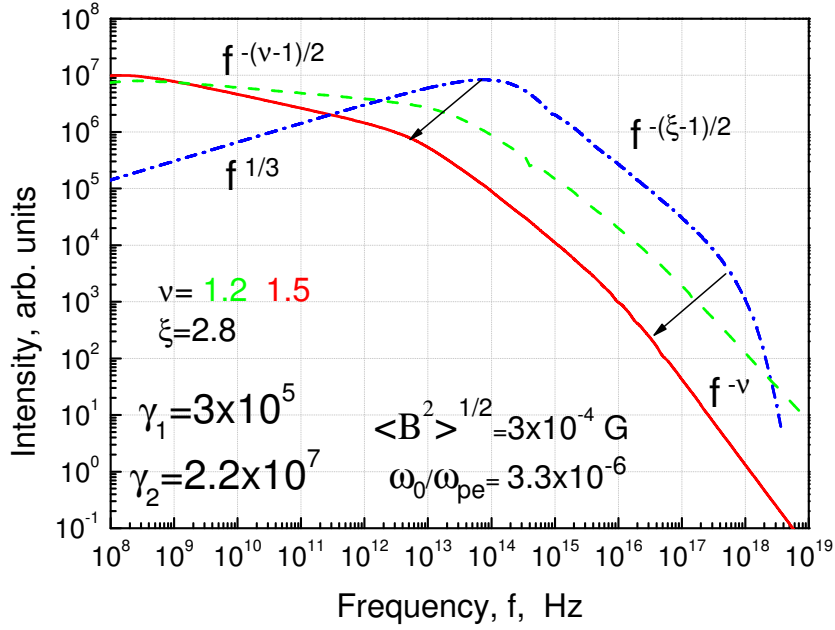} \caption{DSR radiation spectra for a
power-law magnetic field spectrum characterized by different
spectral indices, $\nu$.  The green/dashed curve is $\nu=1.2$, the
red/solid curve is $\nu=1.5$.  In addition, the blue/dash-dotted
curve shows the standard synchrotron spectrum for the same
power-law electron distribution and the same magnetic field energy
density. Arrows indicate the changes of the break frequencies
which separate various spectral asymptotes.  A region with the
standard non-thermal spectrum, $P_\omega \propto
\omega^{-\alpha\nth}$, is present in all cases, although at
differing frequency ranges and at different levels. Beyond this
region, the spectra differ significantly from each other.  As in
Figure~\ref{DSR_Crab}, the values for the energy index of the
power-law electron distribution, $\xi$, the plasma frequency,
$\omega\pe$, and the mean square of the random magnetic field
$\left<B\st^2\right>$ are indicated in the figure.
\label{DSRvsSR}}
\end{figure*}

What are the implications of DSR for PWNe? As has been shown,
differences in the shape of the radiation spectrum between DSR
and standard synchrotron theory will occur when a particular
spectral region is formed primarily by electrons from near either
end of the electron distribution ($\gamma_1$ or $\gamma_2$).  It
is currently believed that a pulsar driving a PWN supplies it with
a wind of highly relativistic electrons, which may have $\gamma_1$
as large as $10^6$ \citep[e.g.,][]{WilsonR1978, KennelC1984a,
KennelC1984b, MelatosM1996, Chevalier_2000, Arons2002}. If this is
the case, and the PWN emission is produced by synchrotron
mechanism, then the broad-band spectrum in the radio, and possibly
up to the infrared, would have the form $P_\omega \propto
\omega^{1/3}$ (see Fig.~\ref{DSRvsSR}), which is not observed.  It
has been suggested that there are in fact two distributions of
electrons: the one mentioned above, called the \emph{nebular}
electrons, and a second distribution to fit the radio
observations, called \emph{radio} electrons.  The postulated radio
electrons have a relatively flat spectrum at $\gamma \ll 10^6$.

In particular in the case of the Crab Nebula, which
is the best studied PWN, a standing shock forms in the pulsar wind
at a distance of $\sim 0.1$~pc from the pulsar.  This shock is
thought to form a power-law energy spectrum of nebular electrons
which produce the bulk of the nebula's synchrotron emission
\citep[see e.g.,][]{KennelC1984a, KennelC1984b, Chevalier_2000,
Arons2002}. Highly mobile features, called wisps, which have been
associated with this shock are observed in optical and X-ray
emission, \citep[e.g.,][]{Hester+1996, Mori+2002}.

The observation of the wisps also in the radio
\citep{BietenholzFH2001, Bietenholz+2004} suggests that the electrons
responsible for the radio emission are accelerated in the same region
as those responsible for the higher-energy emission.  However, if the
radio emission is produced by low ($\gamma < 10^4$) energy ``radio''
electrons, then their origin is unclear, since
it is difficult to produce them in sufficient quantity from a pulsar
wind with high $\gamma$ \citep[e.g.,][]{Arons2002, Atoyan1999}.  It
would, therefore, be highly desirable to interpret the whole PWN
spectrum with the single electron population, rather than requiring a
separate population of radio electrons.

The DSR mechanism indeed suggests a simple and straightforward
interpretation of the observed wide-band PWN spectra with only a
single population of electrons. Within the DSR model, the flat radio
spectrum (with spectral indices $\alpha\rad = 0.2\pm0.2$) should be
associated with the low-frequency (non-perturbative) DSR asymptote
$P_\omega \propto \omega^{-(\nu-1)/2}$.  The spectral index required
of the random magnetic field, therefore, is in the range $\nu=$1 --
1.8, which is in remarkable agreement with current
turbulence models \citep{turbulence, Schek_Cowley_2005}.

The optical emission (and possibly also the millimetre or infrared
and/or to X-ray emission) then, is produced by the inner part of
the electron distribution (equation \ref{distr_fun_power}) and has the
standard form $P_\omega \propto \omega^{-\alpha\nth}$. Finally, at
the frequencies $\omega \gg \omega\lsc \gamma_2^2$, which can
occur in the X-ray or gamma-ray range depending on the actual
value of $\gamma_2$, which is also a decreasing function of the
distance from the pulsar due to significant radiative losses at
these high energies \citep[see, e.g.,][]{BocchinoBykov2001}, the
DSR model predicts a spectrum $P_\omega \propto \omega^{-\nu}$,
resembling the spectrum of relatively small-scale magnetic
inhomogeneities. Interestingly, within the simplest model which
assumes a single power-law spectrum of the random magnetic field
with the index $\nu$, we expect a specific correlation between the
radio spectral index $\alpha\rad=(\nu-1)/2$ and X-ray (or
gamma-ray) spectral index $\alpha\Xray=\nu$:
\begin{equation}
\label{x_r_indices}
   \alpha\rad=(\alpha\Xray-1)/2.
\end{equation}
Note that this equality may not hold exactly if the spectrum of the
magnetic irregularities deviates from a single power-law with
index $\nu$.

Can DSR reproduces the other characteristics of PWNe emission? The
most important such characteristic is probably the significant
polarization often seen in PWNe \citep[e.g.,][]{BietenholzK1991}.
In our simplified case of a completely random field, the average
degree of polarization will evidently be zero if the turbulence is
isotropic. However, modern models of the astrophysical turbulence
\citep{Goldreich_Sridhar_1995, Beresnyak+2005} suggest the
turbulence is highly anisotropic.  In particular, the field
downstream from a strong shock, such as that in a pulsar wind, can
be almost two-dimensional, with virtually all the power in the
plane of the shock.  Such highly anisotropic field seems common in
relativistic shocks, as it has been seen in simulations of a
variety  of shocks \citep[see, e.g., recent reviews by
][]{Piran_2005, Silva_2006}.

The radiation produced in the presence of a field largely
confined to a plane will be highly polarized in the direction
normal to the plane. For example, if the $k$-vectors of the random
waves composing the turbulent magnetic fields line in a plane, the
degree of polarization can be as large as 50$\%$ in the
high-frequency range of the DSR spectrum \citep{Topt_Fl_1987b}.
Furthermore, the presence of a regular component of the field will
also lead to polarized radiation.  A detailed analysis of the
polarization patterns produced in the case of a combination of
regular and random fields applicable to PWNe is beyond the scope
of the current paper.  However, it is clear that a combination of
regular and random fields with anisotropic turbulence can
potentially produce rather strongly polarized radiation.

\section{Comparison to Observed Pulsar Wind Nebulae Spectra}
\label{scompare}
\subsection{The Crab Nebula}
\label{scrab}

The Crab Nebula has been well observed at all wave-bands from the
radio to the gamma-ray.  We compare the broad-band spectrum of the
Crab, taken from \citet{AtoyanA1996} and \citet{AharonianA1998}, with
those obtained using DSR\@.  As was pointed out in \S~\ref{simplic}
above, within conventional synchrotron theory, the combination of a
relatively flat radio spectrum but steeper emission spectra at higher
frequencies is impossible to reconcile with a single power-law energy
spectrum of electrons.

DSR theory, however, can reproduce the entire broad-band spectrum
of the Crab nebula from a single population of nebular
relativistic electrons without introducing a second population of
the radio electrons. A DSR spectrum providing a good fit to the
observed broad-band spectrum of the Crab Nebula is shown in
Figure~\ref{DSR_Crab}. Remarkably, this DSR spectrum was
calculated by using the parameters commonly accepted for the Crab,
i.e., $B\rms=3\cdot 10^{-4}$~G, $\gamma_1=3\cdot 10^{5}$, and
$\xi=2.6$. Compared with the case of the regular field, there are
two new parameters specifying the properties of the random
magnetic field: the largest spatial scale in the random field,
$L_0$, and the corresponding spectral index, $\nu$. These values
adopted for these parameters were $L_0=10^{14}$~cm, which
corresponds to the regime of \emph{large-scale}\/ random magnetic
field, and $\nu=1.54$ to match the observed radio-to-infrared
spectral index. We should note that the DSR spectrum depends on
$\omega\lsc$, which is the combination of $\left<B\st^2\right>$
and $L_0$, equation (\ref{w_lsc_def}), rather than on
$\left<B\st^2\right>$ and $L_0$ separately (as far as $\omega_{0}
\la \omega\pe/\gamma_1$). Therefore, the spectrum will remain
unchanged if the rms magnetic field value scales with $L_0$ as
$B\rms \propto
L_0^\frac{\nu-1}{2}$.

DSR is therefore capable of reproducing the radio spectrum of the Crab
without recourse to a large population of ``radio-emitting'' electrons
with $\gamma \la 10^4$.  As the electrons responsible for the radio
emission are more energetic than in the standard synchrotron theory, a
smaller number of them is required.  The Crab's spectral luminosities
at 1~GHz and 10$^4$~GHz are $\sim 5 \times
10^{24}$~erg~s$^{-1}$~Hz$^{-1}$ and $\sim
10^{24}$~erg~s$^{-1}$~Hz$^{-1}$ respectively (assuming a distance of
2~kpc). The total number of relativistic electrons required to produce
these luminosities can then be calculated from
Equation~\ref{I_DSR_medi}, and is $2 \times 10^{49}$.  Over the
lifetime of the nebula, this corresponds to an average injection rate
of $10^{38.5}$ electrons per second, or, assuming $\gamma \sim 10^6$,
an average energy injection rate of $\sim 10^{38.5}$~erg~s$^{-1}$.
This injection energy is reasonable in light of the pulsar's spindown
luminosity of $\sim 5 \times 10^{38}$~erg~s$^{-1}$, and consistent
with other estimates of the current injection rate
\citep[e.g.,][]{Arons2002}.

\subsection{3C~58}

\begin{figure*}
\includegraphics[height=4in]{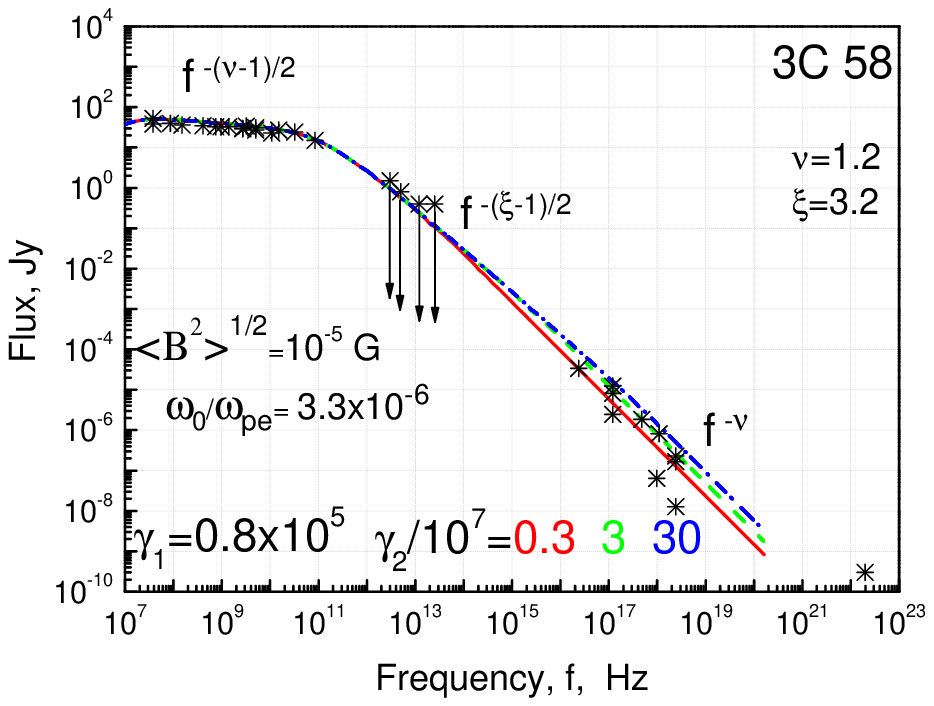}
\caption{The model DSR spectra and observed broad-band spectrum of
3C~58. The radio data are taken from \citet{Green1986},
\citet{GreenS1992}, \citet{MorsiR1987}, and \citet{Salter+1989}.  The
infrared upper limits are from \citet{GreenS1992} and
\citet{Green1994}.  The X-ray data are from \citet{DavelaarSB1986},
\citet{HelfandBW1995}, \citet{BeckerHS1982}, \citet{Torii+2000}, and
\citet{GotthelfHN2006}, with
plotted gamma-ray value being cited in Torii et al. The values adopted
for the energy index of the power-law electron distribution, $\xi$,
the plasma frequency, $\omega\pe$, and the mean square of the random
magnetic field $\left<B\st^2\right>$ are indicated in the figure.
\label{DSR_3C58}}
\end{figure*}

Perhaps the second most intensely studied PWN is 3C~58. We have
compiled measurements of its broadband spectrum extending from a few
MHz in the radio to X-ray, and show the resulting spectrum in
Figure~\ref{DSR_3C58}. There is a significant difference between the
spectrum of 3C~58 and the Crab: In the Crab, the spectrum is an
unbroken power-law from a below 100~MHz to $\sim$15,000~GHz, where a
break with a change in slope of $\Delta\alpha = 0.5$ occurs. This
break is interpreted within standard synchrotron theory as the
synchrotron ageing break \citep[e.g.,][]{Chevalier_2000}. In 3C~58, by
contrast, a spectral break must occur near $\sim 100$~GHz, with a
change in slope which is $> 0.5$ \citep{GreenS1992}.  With standard
synchrotron theory, a break with $\Delta\alpha > 0.5$ cannot be
produced, and furthermore the low frequency of the break in 3C~58
implies an unreasonably large magnetic field \citep[these difficulties
have previously been pointed out
by][]{GreenS1992,Woltjer+1997,Salvati+1998}.

DSR however, naturally produces both spectra with several breaks
and spectral breaks with $\Delta\alpha > 0.5$.  In particular, in
Figure~\ref{DSR_3C58}, the spectrum of 3C~58 is reproduced with the
parameters $B\rms= 10^{-5}$~G, $\gamma_1=0.8\cdot 10^{5}$,
$\xi=3.2$, $L_0=10^{14}$~cm, $\nu=1.2$, not very different from
the parameters adopted for the Crab above. Remarkably, two quite
differently looking spectra, those of Crab and 3C~58, can easily
be reproduced within the same regime of DSR, although with
appropriately adjusted parameter values.

\subsection{Other PWNe}
\label{sothers}

Broad-band spectral measurements as detailed as those for the Crab
Nebula and 3C~58 are not available for other PWNe.  Nevertheless,
the radio and the X-ray spectral indices have been measured for a
good number of systems. Equation (\ref{x_r_indices}) suggests that
within the DSR model there should be a tight correlation between
radio and high-energy (X-ray or gamma-ray) spectral indices of the
emissivity of PWNe. It is known, however, that the size of a
nebula observed in the X-rays is typically smaller than the
corresponding radio and optical size \citep[although
a weak X-ray corona can extend outside the radio emitting
region, see, e.g.,][]{BocchinoBykov2001}.
This means that radiative losses of high-energy electrons are
important and the overall PWN spectrum represents a convolution of
the local emissivity spectrum with actual inhomogeneous spatial
distribution of the emitting material.

As a result, the X-ray spectrum will represent an appropriately
weighted average of two DSR spectral asymptotes, namely $P_\omega
\propto \omega^{-\alpha\nth}$ and $P_\omega \propto
\omega^{-\nu}$, and the X-ray spectral index of the whole nebula
will lie somewhere in between these two values, with the softest
possible X-ray spectral index being $\alpha\Xray=\nu$,
while the hardest one being $\alpha\Xray=\alpha\nth$.
Stated another way, equation (\ref{x_r_indices}) offers a kind of
``line of death'' rather than a strict correlation.

\begin{figure*}
\includegraphics[height=4in]{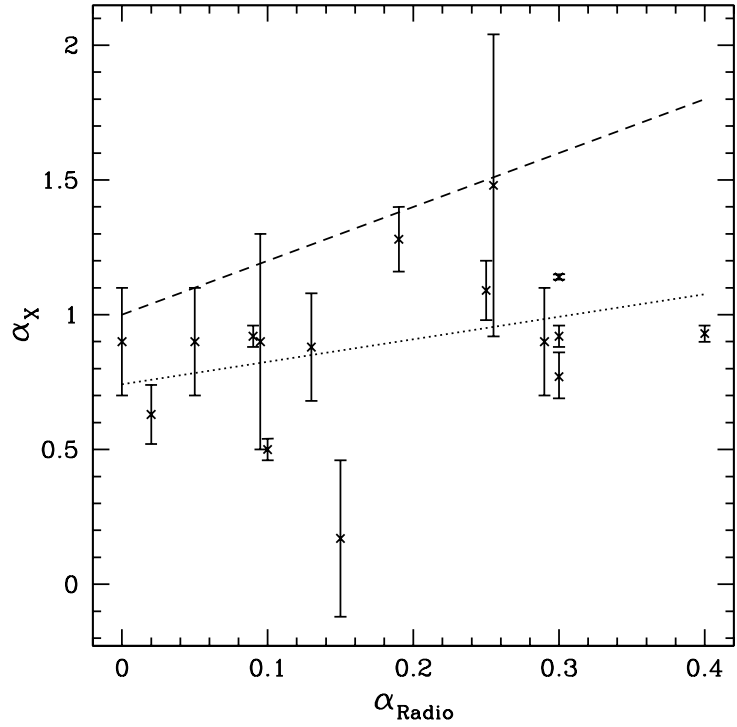}
\caption{A correlation plot of the X-ray ($\alpha_{\rm X}$) vs.\ radio
($\alpha_{\rm Radio}$) spectral indices for the 16 PWNe for which the
spectral indices were available in the literature. The dotted
line is a linear fit: $\alpha\Xray = 0.84\,\alpha\rad + 0.74$, with
the correlation coefficient $R= 0.33$. The dashed line is the ``line
of death'' predicted within the simplest version of the DSR model
based on equation~(\ref{x_r_indices}) as described in
\S\ref{sothers}. The radio and X-ray spectral indices and their
sources are listed in Table~\ref{tspix}. \label{PWNS_spix}}
\end{figure*}

We have compiled from the literature measurements of the X-ray and
radio spectral indices for 16 PWNe.  They are tabulated in
Table~\ref{tspix}, and Figure~\ref{PWNS_spix} displays the radio and
X-ray spectral indices. Note that in all but one cases $\alpha\Xray >
0.5$, which is a bounding value for non-thermal emission from shock
accelerated electrons. Remarkably, all the PWNe obey the line of death
predicted based on the DSR model of PWN emission, which is
clearly supportive of the DSR model, although more data is highly
desirable to increase the statistical significance of this finding.

\section{Discussion and Conclusions}
\label{sconclude}

The radiation produced by a charged particle in the presence of
stochastic magnetic field differs significantly from the standard
synchrotron radiation generated in the presence of large-scale regular
magnetic field only. The corresponding radiative process, called
\emph{diffusive synchrotron radiation} (DSR), can produce a variety of
spectral shapes depending on the parameter combination.

This paper describes a special regime of DSR, namely that produced
in the presence of a strong stochastic magnetic field with
structure on a wide variety of scale, but without any regular
field. The emission spectra in this regime are obtained from the
analysis of the general DSR theory presented in
\citet{Topt_Fl_1987}, \citet{Topt_etal_1987} and \citet{Fl_2005b}.
We emphasize, however, that even though the general theory
includes the special case studied in this paper, the latter has
not been clearly formulated and particularly discussed yet.

Stochastically turbulent magnetic fields are common in astrophysical
objects. In particular, it is widely believed that relatively strong
stochastic component of the magnetic field can be generated by the
highly relativistic wind in PWNe \citep{Kazimura_1998}.
Interestingly, the idea of highly turbulent magnetic field in a PWN
came from observations much earlier: \cite{ReynoldsA1988} interpreted
the presence of sharp filament edges in 3C~58 as a result of highly
diffusive motion of the relativistic electrons due to enhanced Alfv\'en
turbulence; specifically, \cite{ReynoldsA1988} found that $\delta B /
B > 0.1$ on scales of about 10$^{11}$~cm is needed to provide the
required diffusion coefficient.

Accordingly, we applied this specific DSR regime to interpret the
whole broad-band PWN spectra by DSR produced by a single nebular
population of the relativistic electrons without additional radio
electrons. We reached a remarkably good agreement between the
model and observations for two the most studied objects, the Crab
Nebula and 3C~58, and also found a statistical evidence in favour
of this model based on analysis of the correlations between radio
and X-ray spectral indices for a large number of PWNe.

We should mention two shortcomings of simplest version of the DSR
model for PWNe presented here. First, in this paper we assumed
that the random magnetic field is statistically isotropic, which
is typically appropriate for calculating the radiation intensity,
but might be highly incorrect for calculating the polarization of
radiation. As we mentioned in \S \ref{simplic} in the case of
anisotropic stochastic magnetic field the degree of DSR
polarization can be as high as 50$\%$ or more.  There are two
possible avenues of reconcile the DSR model with the high observed
polarization.  First, the magnetic turbulence might be
substantially anisotropic. This conclusion conforms to modern
models of the turbulence generation in astrophysical objects and
fast shocks. Second, we considered only the random component of
the magnetic field and ignored completely any regular component,
which is likely also present in the nebula. Nonetheless, that even
this simplest version of the DSR model gives excellent fit to the
broadband PWN spectra with only a single power-law population of
electrons, which is not possible with the standard synchrotron
model. We believe that this success of the DSR model calls for
developing the DSR theory further to include the joint effect of
strong random and regular fields and to accurately calculate the
degree of polarization in various DSR regimes.

We conclude that the interpretation of the broad-band PWN spectra
using the DSR model is self-consistent and also offers more ways to
observationally study the properties of turbulent magnetic fields such
as those produced by relativistic wind in the PWNe.

\section*{Acknowledgements}

The National Radio Astronomy Observatory is a facility of the
National Science Foundation operated under cooperative agreement
by Associated Universities, Inc. This work was supported in part
by the RFBR grants 06-02-16295 and 06-02-16859. Research at York
University was partly supported by NSERC\@. We have made use of
NASA's Astrophysics Data System Abstract Service. We are very
grateful to D. A. Frail for stimulating our collaboration for this
paper, and to A. M. Bykov for discussions.

\bibliographystyle{mn2e} \bibliography{DSR_PWNs,PWNS-spix}

\begin{table*}
\begin{minipage}{135mm}
\caption{Radio and X-ray Spectral Indices of PWNe
\label{tspix}}
\begin{tabular}{l l l c c l c}
\hline
\multicolumn{2}{c}{Name} &
\colhead{Radio Spectral Index\tablenotemark{a}} &
\colhead{Ref.\tablenotemark{b}} &
~~~~ &
\colhead{X-ray Spectral Index\tablenotemark{c}} &
\colhead{Ref.\tablenotemark{b}} \\
\hline
G16.7+0.1    &        &$ 0.15\pm0.05$ &3& & $0.11\pm0.29$&4  \\
G21.5$-$0.9  &        &  0.0       &1& & $0.90\pm0.2$ &5  \\
G29.7$-$0.3  &(Kes 75)&$ 0.30\pm0.20$ &3& & $0.92\pm0.04$&2 \\
G54.1+0.3    &        &$ 0.13\pm0.05$ &6& & $0.64\pm0.18$&2  \\
G74.9+1.2   &(CTB87)  &$ 0.26$        &7& & $1.48\pm0.56$&8  \\
G130.7+3.1  &(3C58)  &$ 0.10\pm0.02$  &9& & $0.92\pm0.04$&2  \\
G184.6$-$5.8&(Crab)  &$ 0.30\pm0.04$ &10& &$1.14\pm0.01$&2  \\
G189.1+3.0  &(IC443) &$ 0.02\pm0.10$ &11& & $0.7\phn\pm0.10$&11  \\
G263.9$-$3.3&(Vela X)&$ 0.10 $       &12& & $0.50\pm0.04$&2  \\
G291.0$-$0.1&(MSH 11$-$62)&
                      $ 0.29\pm0.05$&13& & $0.9\phn\pm0.2$ &13 \\
G292.0+1.8  &(MSH 11$-$54)&$0.05\pm0.05$&14& & $0.90\pm0.20$ &14 \\
G320.4$-$1.2&(MSH 15$-$52)& 0.4    &1 & & $0.93\pm0.03$ &2 \\
G343.1$-$2.3&            &  0.3      &15& & $0.77\pm0.08$ &15 \\
0540$-$69   &            &$ 0.25\pm0.1$&16& & $1.09\pm0.11$ &2\\
N157B    &               &$ 0.19   $   &17& & $1.28\pm0.12$ &2\\
B0543-0685 &             &$ 0.10\pm0.05$&18& & $0.90\pm0.4$ &18 \\
\hline
\end{tabular}

\medskip

\tablenotetext{a}{The radio spectral index ($S \propto
f^{-\alpha}$) of the remnant or of its pulsar wind nebula
segment.}\\

\tablenotetext{b}{References: 1. \citet{Green2004}, see also Green
D.A., 2004, `A Catalogue of Galactic Supernova Remnants (2004
January version)', Mullard Radio Astronomy Observatory, Cavendish
Laboratory, Cambridge, United Kingdom (available at
http://www.\-mrao.\-cam.\-ac.uk/\-surveys/snrs);
2.  \citet{Gotthelf2003};  3. \citet{BockG2005};
4.  \citet{HelfandAG2003}; 5. \citet{MathesonS2005};
6.  \citet{VelusamyB1988}; 7. \citet{MorsiR1987};
8.  \citet{AsaokaK1990};
9.  \citet{Green1986};    10. \citet{Bietenholz+1997};
11. \citet{Olbert+2001};  12. \citet{Hales+2004};
13. \citet{HarrusHS1998};
14. \citet{GaenslerW2003}; 15. \citet{Romani+2005};
16. \citet{ManchesterSK1993}; 17. \citet{Lazendic+2000};
18. \citet{Gaensler+2003}}\\

\tablenotetext{c}{The X-ray spectral index ($S \propto
f^{-\alpha}$) of the remnant or of its pulsar wind nebula
segment.}\\
\end{minipage}
\end{table*}

\clearpage

\end{document}